\documentclass[a4paper,12pt]{article}
\usepackage{amsmath, amssymb, amsthm}
\usepackage{fullpage}
\usepackage{hyperref}
\usepackage{graphicx}
\usepackage{parskip}
\usepackage{color}
\usepackage{appendix}
\usepackage{multicol}
\usepackage{comment}

\newcommand*{\rom}[1]{\expandafter\@slowromancap\romannumeral #1@}

\numberwithin{equation}{section}

\newcommand{\figref}[1]{Figure \ref{#1}}

\renewcommand{\rho}{|\psi|}

\newcommand{\C}{\mathbb{C}}
\newcommand{\Z}{\mathbb{Z}}

\graphicspath{{Images/}}

\begin{document}
\title{Type 1.5 vortex bags in single component Ginzburg-Landau models with higher order potentials}
\author{Martin Speight$^*$ and Thomas Winyard$^\dagger$ \\[10pt]
{\em \normalsize $^*$School of Mathematics, }\\{\em \normalsize University of Leeds, Leeds, LS2 9JT, U.K.}\\[10pt]
{\em \normalsize $^\dagger$Division of Mathematics,}\\{\em \normalsize  University of Dundee, Dundee, DD1 4HN, U.K.}\\[10pt]
{\normalsize $^*$j.m.speight@leeds.ac.uk \quad  $^\dagger$twinyard001@dundee.ac.uk}}

\maketitle
\begin{abstract}
We demonstrate the existence of type 1.5 like superconductivity in a single component Ginzburg-Landau model with a higher order potential. The model exhibits two competing superconducting Meissner states (MS), each of which has its own pair of length scales (coherence length and penetration depth). The solutions take the form of bags that contain vortices of one vacuum inside the other. This gives rise to four different types of vortex interaction, where one set of length scales determines vortex interactions inside the bags and the other set between the bags. Hence, within the bag the vortices can repel or attract and the bags themselves can repel or attract depending on the hierarchy of four length scales. When vortices repel within the bag, solutions act similarly to those in type 1.5 multicomponent superconductors, but in a single component system.
\end{abstract}

\vspace{15pt}

\section{Introduction}

In the simplest case a superconductor can be well described by an effective model of a single complex order parameter field $\psi$, coupled to a $U(1)$ gauge field $A$. This effective field theory, coined the Ginzburg-Landau model, exhibits two fundamental length scales, the magnetic field penetration depth $\lambda$ and the characteristic length scale associated with the order parameter, the coherence length $\xi$. Ginzburg and Landau \cite{landau1950k} demonstrated that their ratio $\kappa = \lambda/\xi$ determines the response of a superconductor to an external magnetic field, sorting  them into two categories: type I when $\kappa < 1$ (coherence length longer) and type II when $\kappa > 1$ (penetration depth longer). A key feature of vortices in type I (type II) superconductors is that they exhibit attractive (repulsive) intervortex forces at all distances.

In recent decades, there has been increased interest in unconventional superconductors. These are described by multicomponent systems, requiring $n$ (often highly coupled) complex fields $\psi_i$, $i \in \{1,2,\ldots,n\}$. These unconventional models exhibit $n$ coherence lengths, and a single parameter can no longer describe their response to applied fields. The type~I (type II) regime persists when all the coherence lengths are longer (shorter) than the penetration depth. However, in the case where the penetration depth lies between the smallest and largest coherence length a new mixed state exists \cite{babaev2005semi,Babaev:2010type}, coined type 1.5 superconductivity \cite{moshchalkov}.

This type 1.5 state exhibits vortex interactions that are attractive at long range but repulsive at short range, leading to multi-vortex bound states or clusters. As such non-monotonic vortex interactions require multiple length scales, it appears on the surface that multiple components are required for such vortex clusters to appear in the bulk of a superconductor. Such materials are rare but there has been recent evidence of such vortex clusters occurring in materials such as ${\rm Sr_2Ru O_4}$ \cite{moler2,Ray.Gibbs.ea:14}, ${\rm LaPt_3 Si}$ \cite{noncentr,1347-4065-54-4-048001}, ${\rm UTe_2}$ \cite{wang2025observation}
and ${\rm Zr B_{12}}$ \cite{biswas2020coexistence,ge2022experimental}.

Recently certain exotic systems have been demonstrated to exhibit these indicative long-range attraction short-range repulsion intervortex forces, of a type 1.5 superconductor, in a single component model. For example, the introduction of higher order gradient terms can produce FFLO systems that can stabilize vortex bound states \cite{barkman2020ring}. In addition, coupling the GL model to a chiral magnet can produce similar vortex bound states \cite{leask2026interactions}. Such exotic systems introduce additional length scales through complicated couplings and are yet to be observed in materials. In contrast, this paper demonstrates a simpler possibility. By expanding the standard GL potential to higher order, we demonstrate the existence of multiple superconducting Meissner states (MS), each of which has its own length scales. This directly leads to type 1.5 behaviour and vortex clustering.

There is already observation of multiple competing superconducting states in materials such as ${\rm CeRh_2As_2}$ \cite{khim2021field}, ${\rm UTe_2}$\cite{wu2024enhanced} and ${\rm UPt_3}$ \cite{joynt2002superconducting}. In such systems interesting behaviour is often observed when two superconducting states become almost degenerate in energy and are likely to coexist \cite{wu2026discovery}. In this paper we study the simplest model of two coexistent superconducting states, demonstrating that this has a significant effect on vortex interactions, even for single component systems.

Finally, the Ginzburg Landau model with higher order potentials has recently become interesting in cosmological models \cite{martinez2026oscillons}, where vortex scatterings have been shown to exhibit resonant structures and oscillons. In this field the effect of mode excitation on the scattering of vortices has become of interest in the study of cosmic strings \cite{krusch2024scattering,alonso2024spectral,krusch2026type,alonso2026bps}

When deriving the Ginzburg–Landau model from microscopic theories \cite{Gorkov}, the free-energy expansion is typically truncated at low order in the superconducting order parameter, neglecting higher-order terms. In this paper we demonstrate that when included, if the higher order terms are relevant, they can induce markedly different vortex interactions. In particular a higher order polynomial in the order parameter magnitude $|\psi|$ can have multiple minima. Each minimum corresponds to a competing Meissner state with its own coherence length $\xi_i$ (set by the curvature of the potential at the minimum) and its own penetration depth $\lambda_i$ (set by the magnitude of the order parameter at the minimum), and hence its own GL parameter 
$\kappa_i=\lambda_i/\xi_i$. If the inner MS has $\kappa_1>1$ while the outer MS has $\kappa_2<1$ then we induce long range attraction and short range repulsion between vortices, analogous to a type 1.5 superconductor. Note that one may also have $\kappa_1<1<\kappa_2$, a new regime that is not analogous to a multicomponent type 1.5 system.

\section{The model}
The single component Ginzburg-Landau (GL) model is a 2-dimensional classical effective field theory, describing a conventional superconductor near the critical temperature $T_c$. It is formulated in terms of a single complex order parameter $\psi:\mathbb{R}^2\rightarrow\mathbb{C}$, which minimizes the free energy,
\begin{equation}
F = \int_{\mathbb{R}^2}\left\{\frac{1}{2}\overline{D_i \psi }D_i \psi + \frac{1}{2} B^2 + V(\psi)\right\},
\label{eq:energy}
\end{equation}
where $A_i$ is the $U(1)$ electromagnetic gauge potential, $D_i = \partial_i - iA_i$ is the corresponding covariant derivative
and $B=\partial_1A_2-\partial_2 A_1$ is the magnetic field. Note that the units of $A$ are chosen to absorb any multiplicative constant, usually written $2 \pi/\psi_0$ (where $\psi_0=h/2|e|$ is the flux-quantum). Physically we are interested in studying the gauge invariant quantities $B$ and $|\psi|$, the magnitude of the order parameter, representing the density of Cooper paired electrons. The resulting equations of motion are
\begin{align}
-D_iD_i \psi + 2 \frac{\partial V}{\partial \psi^*}&=0,\\
-\partial_i(\partial_i A_j - \partial_j A_i)+ \mbox{Im}(\psi^* D_j \psi)&=0.
\label{eq:eom}
\end{align}

The potential term $V$ determines the mass of the order parameter and must be gauge invariant, hence it can only depend on $|\psi|$. The standard GL model uses the minimal expansion for $V$ to quartic order in $|\psi|$, giving the familiar Higgs symmetry breaking potential term,
\begin{equation}
V_{GL}(\psi) = \frac{\mu^2}{8} \left( 1 - \rho^2\right)^2,
\label{eq:Higgs}
\end{equation}
with vacuum value $|\psi| = 1$. The vacuum state has a $U(1)$ symmetry (due to the phase of the complex field $\varphi$).

To ensure a given configuration has finite energy $|\psi(x)|\rightarrow 1$ as $|x| \rightarrow \infty$, but $\psi_\infty : \partial \mathbb{R}^2 \equiv S^1_\infty \rightarrow S^1\subset\C$, may have non-zero winding number $N\in\Z$.  Finite energy  also requires $D\psi \rightarrow 0$ on the boundary, and so, by a standard application of Stokes's theorem we have,
\begin{equation}
\int_{\mathbb{R}^2} B = 2\pi N.
\end{equation}
So the winding number $N$ measure the total magnetic flux through the plane, which is quantized.

The single vortex solution ($N=1$) has, in suitable gauge, radial symmetry around a single zero of $\psi$. For $N>1$ the solutions are determined by the parameter $\mu$. When $\mu < 1$ ($\mu > 1$) the vortices exhibit monotonic attractive (repulsive) interactions and are coined type I (type II) vortices. Hence, in the type I case vortices converge into a single mega-vortex where the phase of $\psi$ winds $N$ times around a single zero of multiplicity $N$. By contrast there are no stable static solutions for $N>1$ for type II as the vortices repel to infinite separation.

In this paper we include higher order terms in the potential, the simplest example of which is,
\begin{equation}
V_1(\psi) = \frac{\mu^2}{8}\left(m_1^2 - |\psi|^2\right)^2\left(m_2^2 - |\psi|^2\right)^2.
\label{eq:V1}
\end{equation}
We may assume, without loss of generality, that $m_1 < m_2$. We now have two degenerate vacua up to phase $\rho = m_a$, so the the vacuum manifold is the disjoint union of two circles $S^1\sqcup S^1$. Thus we now have two species of vortex depending on the value of $\rho$ on the boundary.

Such a vacuum manifold should be interpreted as two competing superconducting Meissner states $\rho = m_1$ or $\rho=m_2$ (each with their own vortex species) and introduces a domain wall that interpolates between the two disconnected circles.

Interactions between vortices will depend on the mass of the field $\psi$ about the chosen vacuum and hence how steep $V$ is about $m_a$. This means the field mass and hence the usual classification into type~I or type~II depend on the vacuum.

Above, we have presented a potential that has degenerate energy which is rarely the case for competing superconducting states (which often change energy hierarchy based on the parameters of the system). In addition, we want to be able to control how steep the potential is about each vacuum. Hence, we will generalize the potential to the following,
 \begin{equation}
V(\psi) = \frac{\mu^2}{8}\left[\left(m_1^2 - |\psi|^2\right)^2\left(m_2^2 - |\psi|^2\right)^2\left(1 + |\psi|^2\right)^\alpha + \beta |\psi|^2\right],
\end{equation}
where we recover $V_1$ if $\alpha = \beta = 0$. We will use this potential for the rest of the paper, giving us three parameters $(\mu,\alpha,\beta)$ that control the mass of the order parameter at the two competing Meissner states $\rho =u_1\approx m_1$ and $\rho= u_2\approx m_2$, and which Meissner state has lower energy. Note that $\beta > 0$ ($\beta < 0$) means $u_1$ ($u_2$) is the true ground state, while the other is only a locally stable superconducting state. In addition, $\alpha > 0$ ($\alpha < 0$) increases (decreases) the mass of the field at both vacua but affects $u_2$ to a greater degree.

\section{Vortex asymptotics and long range interactions}

\begin{figure}
  \centerline{\includegraphics[width=\linewidth]{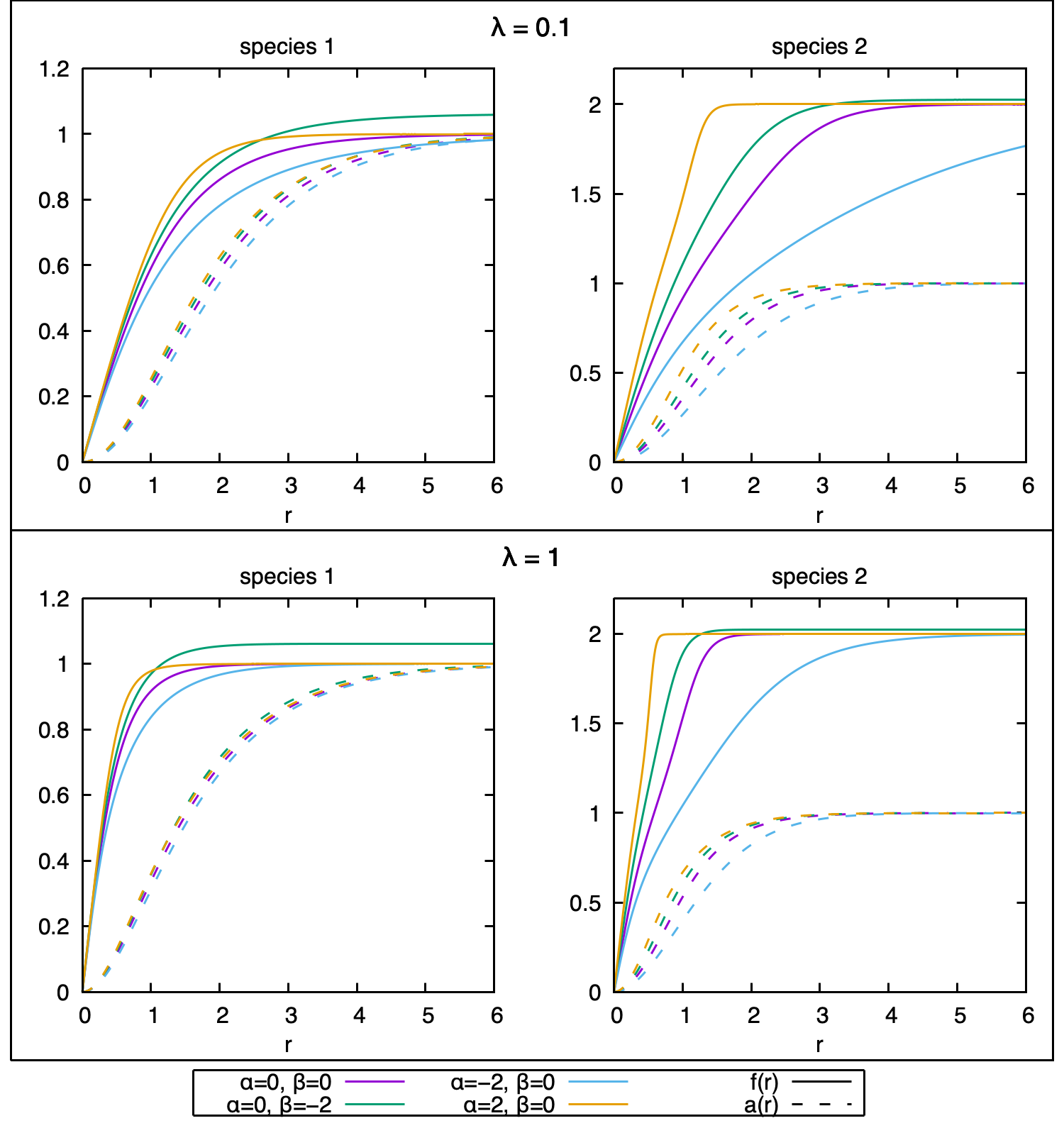}}
  \label{fig:degree1}
 \caption{Profile solutions to \eqref{eq:radial_eom} for a single vortex ($N=1$).}
\end{figure}

We assume that the single vortex ($N=1$) solution retains the $O(2)$ rotational symmetry of the energy in \eqref{eq:energy} (confirmed by numerics). Thus we consider fields of the form,
\begin{align}
\psi &= f(r) e^{i\theta},&
\left(A_1, A_2\right) &= \frac{a(r)}{r}\left(- \sin{\theta}, \cos{\theta}\right),
\end{align} 
where $r,\theta$ are polar coordinates in the plane and $f$ and $a$ are real profile functions. Regularity requires that at the origin $f(0) = a(0) = 0$ while finite energy requires that,
\begin{align}
\lim_{r\rightarrow \infty} f(r) &= u_a,& 
\lim_{r \rightarrow \infty} a(r) &= 1.
\end{align}
The choice of $u_a$ determines which of the two species of vortex we are considering. Substituting the above ansatz into the equations of motion \eqref{eq:eom} leads to the system of ordinary diferential equations
\begin{align}
f'' + \frac{1}{r} f' - \frac{1}{r^2}(1 - a)^2 f &= V'(f),\\
a'' - \frac{1}{r} a' + (1 - a)f^2 &= 0.
\label{eq:radial_eom}
\end{align}
These field equations are non-linear and must be solved numerically, the results of which are plotted for a number of parameters in \figref{fig:degree1}. Note that the profile for species 1 ($u = u_1$) remain similar to those from standard GL theory, however when the outer superconducting state $u_2$ is chosen at infinity the profiles change form. This is unsurprising as for species 2 the potential $V(f)$ is no longer monotonic on the range $f \in [0,u_2]$.

While we cannot find analytic solutions to \eqref{eq:radial_eom} we can study the fields far from the vortex core that dominate the long-range interactions. Hence, we introduce the small quantities $\widehat{f} := f - u_a$, $\widehat{a} := a - 1$ and expand the field equations to linear order in $\widehat{f}$ and $\widehat{a}$, such that
\begin{align}
\widehat{f}'' + \frac{1}{r} \widehat{f}' &= \left. \frac{\partial^2 V}{\partial |\psi|^2}\right|_{(u_a, 0)}\widehat{f},\\
\widehat{a}'' - \frac{1}{r}\widehat{a}' - u_a^2 \widehat{a} &= 0.
\end{align}
These are simply decoupled modified Bessel's equations with solutions,
\begin{equation}
\widehat{a} = \frac{q^a_0}{2\pi} r K_1(u_a r), \quad \quad \widehat{f} = \frac{q^a_1}{2\pi} K_0(\mu_a r),
\end{equation}
where $K_n$ denotes the $n$th modified Bessel's function of the second kind with the length scales,
\begin{equation}
\lambda_a = \frac{1}{u_a}, \quad \quad 
\xi_a = \frac{1}{\mu_a} = V''(u_a)^{-\frac{1}{2}}
\end{equation}
Here $\lambda_a$ and $\xi_a$ can be interpreted as the London penetration depth and coherence length respectively> Note that we have two sets of length scales depending on which superconducting state $a=1,\, 2$ we are linearizing about.

We can now apply the familiar typology argument that is made in standard GL models. Here the dominant length scale at long range is determined by which length scale is larger, then type I (type II) indicates that $\xi_a > \lambda_a$ ($\lambda_a > \xi_a$). Hence, the key observation of this paper is that, for chosen parameters we have two different sets of length scales and hence two different superconducting states of different types in general. That is, all four of the following regimes
are possible:
\begin{itemize}
\item type I/I - both superconducting states are type I.
\item type II/II - both superconducting states are type II.
\item type II/I - the inner superconducting state $u_1$ is type II while the outer state $u_2$ is type I ($u_1 < u_2$).
\item type I/II - the inner superconducting state $u_1$ is type I while the outer state $u_2$ is type II ($u_1 < u_2$).
\end{itemize}

If we now restrict to specific families of parameters we can understand the typology for different systems. First consider the simplest case $\alpha = \beta = 0$ in which the potential is of the form \eqref{eq:V1} with two degenerate vacuum states $u_a = m_a$. Then
\begin{equation}
V''(u_a) = \mu^2 m_a^2 (m_1^2 - m_2^2)^2,
\end{equation}
and so the length scales are,
\begin{equation}
\lambda_a = \frac{1}{m_a},\quad\quad
\xi_a = \frac{1}{m_a}\frac{1}{\mu|m_1^2 - m_2^2|}.
\end{equation}
Hence the ratio of the length scales does not change between superconducting states,
\begin{equation}
\kappa = \frac{\lambda_a}{\xi_a} = \mu |m_1^2 - m_2^2|.
\end{equation}
Hence, for this subfamily of potentials we can only realize type I/I ($\kappa < 1$) or type II/II ($\kappa > 1$) behaviour. Other typologies require more control over how steep the potential is about each vacua and hence the mass of the field in that superconducting state.

\begin{figure}
  \centerline{\includegraphics[width=\linewidth]{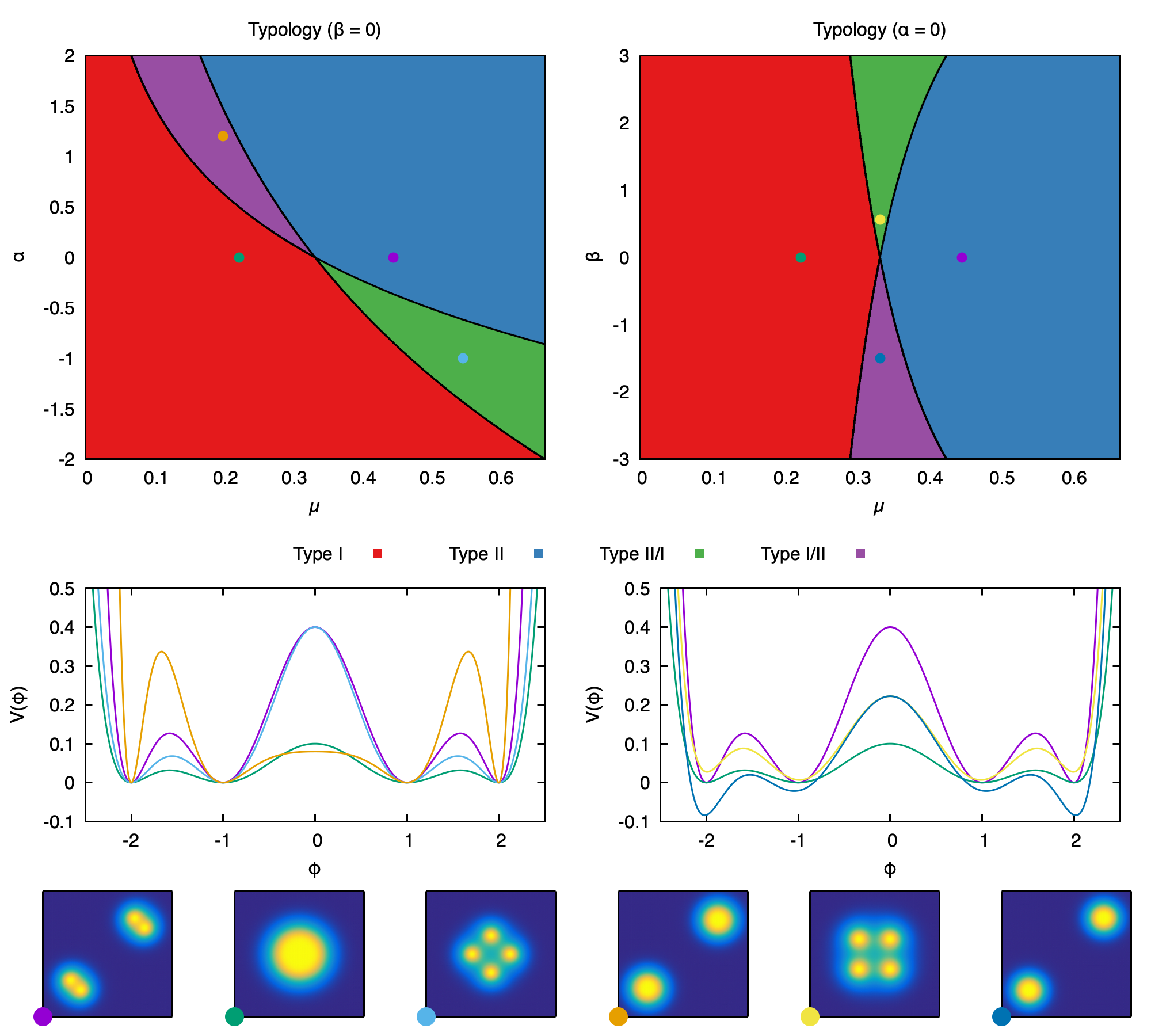}}
  \label{fig:typology}
 \caption{The top row indicates the typology of the model for different parameters. The coloured points indicate the parameters of the potentials plotted in the same colour in row 2 and the vortex configurations in the final row. The configurations are plots of the magnetic field density $B$ for a winding $4$ local minimizer that resulted from an initial condition of two well separated degree $2$ configurations. The typology then determines the interaction type as described in section 3.}
\end{figure}

We can achieve this while retaining two degenerate vacua by considering the family $\alpha \neq 0$, $\beta = 0$. Here we still have $u_a = m_a$, but now
\begin{equation}
V''(u_a) = 
\mu^2 m_a^2 (m_1^2 - m_2^2)^2(1 + m_a^2)^\alpha
\end{equation}
so the ratio of length scales
\begin{equation}
\kappa_a = \mu |m_1^2 - m_2^2|(1 + m_a^2)^{\alpha/2}
\label{eq:kap_1}
\end{equation}
differs between $a=1$ and $a=2$
Hence, we can now have two different types for the different superconducting states. If $\alpha > 0$ then $\kappa_1 < \kappa_2$ and we can have type I/II and alternatively if $\alpha < 0$ then $\kappa_1 > \kappa_2$ such that we can have type II/I. This can be seen in the left panel of \figref{fig:typology}. The curves that separate the different regions are simply calculated by rearranging \eqref{eq:kap_1} when $\kappa_a = 1$ to get the separatrix,
\begin{equation}
\alpha_a(\mu) = -2\frac{\ln \mu + \ln|m_1^2 - m_2^2|}{\ln(1+m_a^2)},
\end{equation}
giving two log curves that meet at the critical point $\mu = 1/|m_1^2-m_2^2|$, $\alpha = 0$. Then the type II/I region is between these curves for $\alpha < 0$ and the type I/II region is between these curves for $\alpha > 0$, as shown in \figref{fig:typology}.

If we now consider the case where $\alpha = 0$ and vary $\beta$, then the two vacuum states $u_a$ are non-degenerate in energy. We follow the same approach as before to find the length scales, which first requires solving $V'(u_a)=0$ for $u_a$. For $u_a\neq0$ this is a cubic in $u_a^2$, which when rearranged gives,
\begin{equation}
\beta=2(m_1^2-u_a^2)(m_2^2-u_a^2)
(m_1^2+m_2^2-2u_a^2).
\label{eq:betaEq}
\end{equation}
While we don't have an explicit form for $u_a$, we can still make progress. As $V$ is a function of $y = |\psi|^2$, we write $V(|\psi|) = W(y)$, such that,
\begin{equation}
V''(|\psi|) = 2W'(y) + 4|\psi|^2\, W''(y).
\end{equation}
Hence, as $u_a$ is a minimum we have $V'(u_a) = 2 u_a W'(u_a^2) = 0$ and $W'(u_a^2) = 0$ for $u_a \neq 0$, such that,
\begin{align}
\nonumber V''(u_a) &= 4u_a^2 W''(u_a^2),\\ 
&= u_a^2\mu^2 \left[ (m_1^2-u_a^2)^2 +(m_2^2-u_a^2)^2 + 4(m_1^2-u_a^2)(m_2^2-u_a^2) \right].
\end{align}

The ratio of the length scales is then given by,
\begin{equation}
\kappa^2_a = V''(u_a)/u_a^2,
\end{equation}
such that the separatrix is found by solving $\kappa_a^2=1$, which is a quadratic in $u_a^2$ with solutions, 
\begin{equation}
u_a^2 = \frac{1}{2}\left[m_1^2+m_2^2 \mp \sqrt{\frac{(m_2^2-m_1^2)^2}{3}+\frac{2}{3\mu^2}}\,\right],
\end{equation}
where the $-/+$ correspond to the smaller/larger vacuum state. Finally, substituting these into \eqref{eq:betaEq} gives the two curves,
\begin{align}
\beta(\mu) = \pm\frac{4}{3}
\sqrt{ \frac{(m_2^2-m_1^2)^2}{12} + \frac{1}{6\mu^2} }
\left[
\frac{1}{2\mu^2} - \frac{(m_2^2-m_1^2)^2}{2}
\right],
\end{align}
where the upper (lower) sign corresponds to the smaller (larger) vacuum. Again the two curves meet at the critical point $\mu = 1/|m_1^2 - m_2^2|$, $\beta = 0$.

For $\beta > 0$ the separatrix for the smaller Meissner state $u_1$ is to the left (smaller $\mu$) of the separatrix for $u_2$ and visa-versa for $\beta < 0$. Hence, for $\beta > 0$ the Type II/I regime is between the two curves while for $\beta < 0$ the Type I/II is between the two curves, the result of which is shown in \figref{fig:typology}.

The most interesting consequence of different typologies for this paper are the static inter-vortex forces. The gauge field mediates a repulsive force between vortices, while the order parameter mediates an attractive force.

We can write an explicit formula for the resulting long-range inter-vortex interaction potential using the point vortex formalism \cite{speight1997static} that models vortices as point particles in a linear theory of a real Klein-Gordon field $\widehat{\psi}$ of mass $\xi_a^{-1}$ and vector potential $\widehat{A}$ of mass $\lambda_a^{-1}$. As a point particle (vortex) then carries scalar monopole charge $q^a_1$ and magnetic dipole moment $q^a_0$ then the interaction potential at seperation $r$ becomes,
\begin{equation}
V_a(r) = (q^a_0)^2 K_0(r/\xi_a) - (q^a_1)^2 K_0(r/\lambda_a).
\end{equation}
Hence, the long range inter-vortex forces will depend on the superconducting state $a\in\{1,2\}$.

Note that while inter-vortex forces are monotonic in the standard GL model, this does not follow from the linearization above which only describes long-range interactions. Short range inter-vortex forces are polynomial scaling as $r^4$ \cite{speight2025short}, so we can only discover if vortices exhibit short-range repulsion (attraction) and hence locally split (coalesce) by non-linear numerical simulations.

\section[Static solutions for N > 1]{Static solutions for $N > 1$}
We now turn to studying vortices for higher winding number $N$. As we can no longer assume rotational symmetry we use a gradient decent approach to minimize a discrete approximation of the energy functional.

Our algorithm is explained in detail in \cite{speight2020skyrmions}: we minimize $F$ in \eqref{eq:energy} over all fields ($\psi$, $A$) on a large rectangle, with fixed boundary conditions such that $\psi$ has winding $N$ on the boundary. We achieve this by using Newton flow, taking the force to be a lattice approximation of minus the $L^2$ gradient of $F$ or, more concretely, minus the left hand side of the equations of motion \eqref{eq:eom}. We use an arresting criterion that sets the velocity of the fields to 0 if the flow starts to move opposite to the direction of the gradient \cite{speight2020skyrmions}. The derivatives were approximated using a fourth order central finite difference scheme.

With two vacua $u_1$ and $u_2$ ($u_1 < u_2$), we have two possible boundary conditions. If we choose $|\psi_\infty| = u_1$ then the solutions for $N>2$ follow the predictions from the typology, where the inner type is type I (II) leads to monotonic attractive (repulsive) interactions. This matches the behaviour of standard single component Ginzburg-Landau with the Higgs potential in \eqref{eq:Higgs}. Hence, we have not plotted these results but focused on the novel case of the boundary condition $|\psi_\infty| = u_2$.

\begin{figure}
  \centerline{\includegraphics[width=\linewidth]{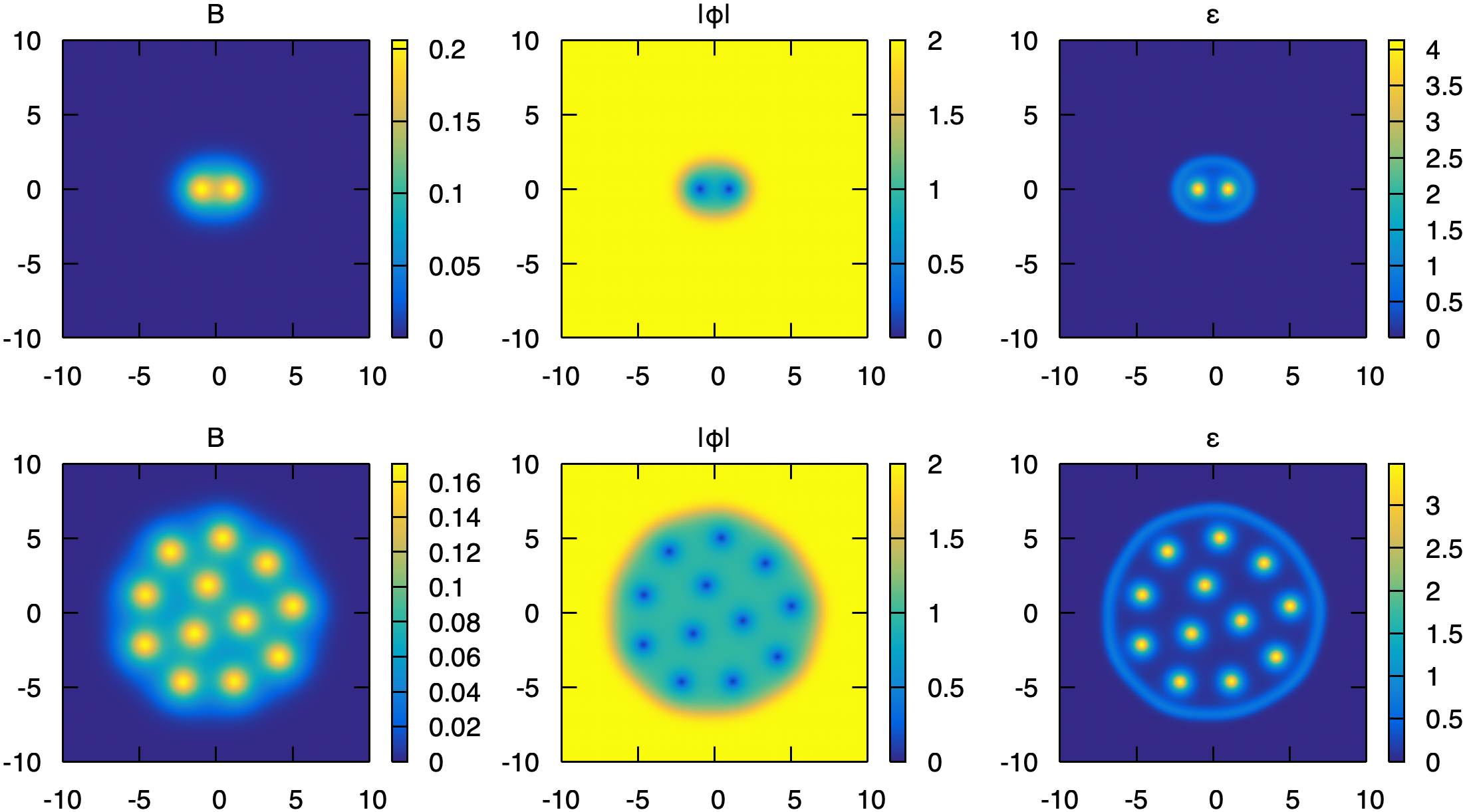}}
  \label{fig:vortexII}
 \caption{Local minimizers of degree $N=2$ and $N=12$ respectively for $\mu^2 = 2$, $m_1 = 1$, $m_2 = 2$ and $\alpha = \beta = 0$. The plots include the magnetic field density $B$ the order parameter magnitude $|\psi|$ and the energy density $\mathcal{E}$. These parameters correspond to type II for both vacua, as can be seen in \figref{fig:typology}, hence they split locally, but are contained by the domain wall into the superconducting bubble.}
\end{figure}

For simplicity all results are presented for the choice $m_1 = 1$ and $m_2 = 2$. The simplest result can be seen in the first row of \figref{fig:vortexII} where we have found a degree 2 ($N=2$) solution for $\mu^2 = 2$ and $\alpha = \beta = 0$. These parameters are in the type II regime and the initial condition took the form of a degree 2 configuration with $O(2)$ symmetry. The interactions naturally split the zeros which repel. The split vortices can be thought of as being in a bubble or bag of the superconducting state $u_1$ which is surrounded by a domain wall which interpolates between $u_1$ and $u_2$. The tension of this domain wall limits the separation of the two vortices. Hence the final local minimizer takes the form of a bound state of $N$ vortices in a bag as shown in \figref{fig:vortexII}.

Note that the domain wall or bag can be clearly seen in the energy density plot in \figref{fig:vortexII}, but not in the magnetic field $B$ plot. This is a result of the domain wall not coupling to the magnetic field, which requires anisotropy \cite{benfenati2020magnetic}. This means in an experiment the domain wall is not readily apparent and all configurations will look identical to type 1.5 vortex clustering.

\begin{figure}
  \centerline{\includegraphics[width=\linewidth]{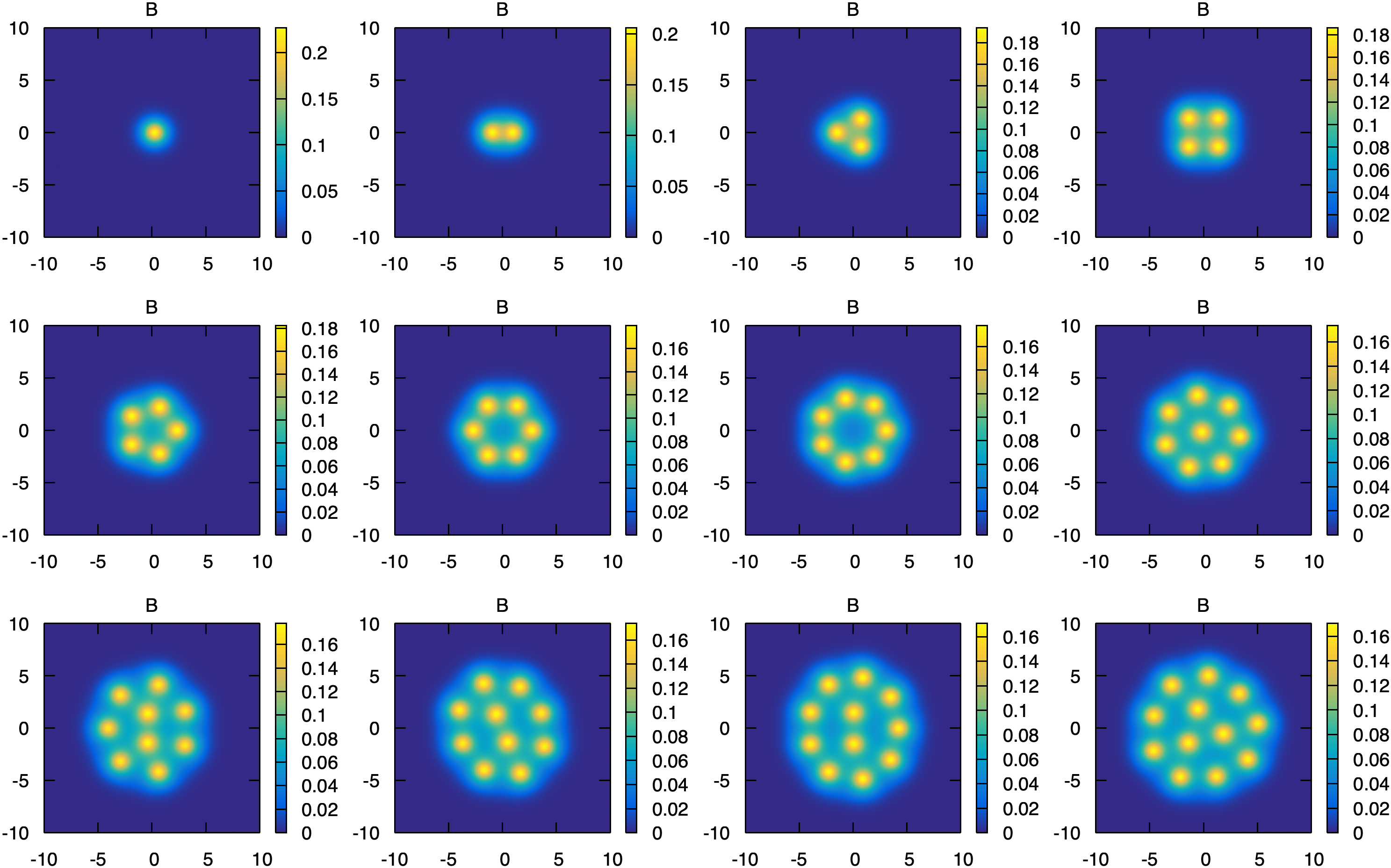}}
  \label{fig:allVortices}
 \caption{Plots of the magnetic field density $B$ for local minimizers of degree $N=1-12$ for parameters $\mu^2 = 2$, $m_1 = 1$, $m_2 = 2$ and $\alpha = \beta = 0$ which is type II.}
\end{figure}

A similar process occurs for higher winding which can be seen in the bottom row of \figref{fig:vortexII} and the resulting magnetic field $B$ for $N = 1-12$ in \figref{fig:allVortices}. As $N$ increases the competition between the repulsion of the vortices and the domain wall tension causes the bag to grow in size. The final vortex positions are similar to those that occur for type 1.5 clusters in multicomponent systems \cite{babaev2017type}. One key difference is that the clusters are always symmetric here. Due to the potential from the domain wall energy, the solitons position themselves similarly to in a packing problem. This is a common feature of solitons in a potential \cite{elliot2015baby,winyard2016skyrme}. Note that for \figref{fig:allVortices} we observe that the maximum magnetic field $B$ decreases with higher winding number $N$ as the vortices separate in the bag.

\begin{figure}
  \centerline{\includegraphics[width=\linewidth]{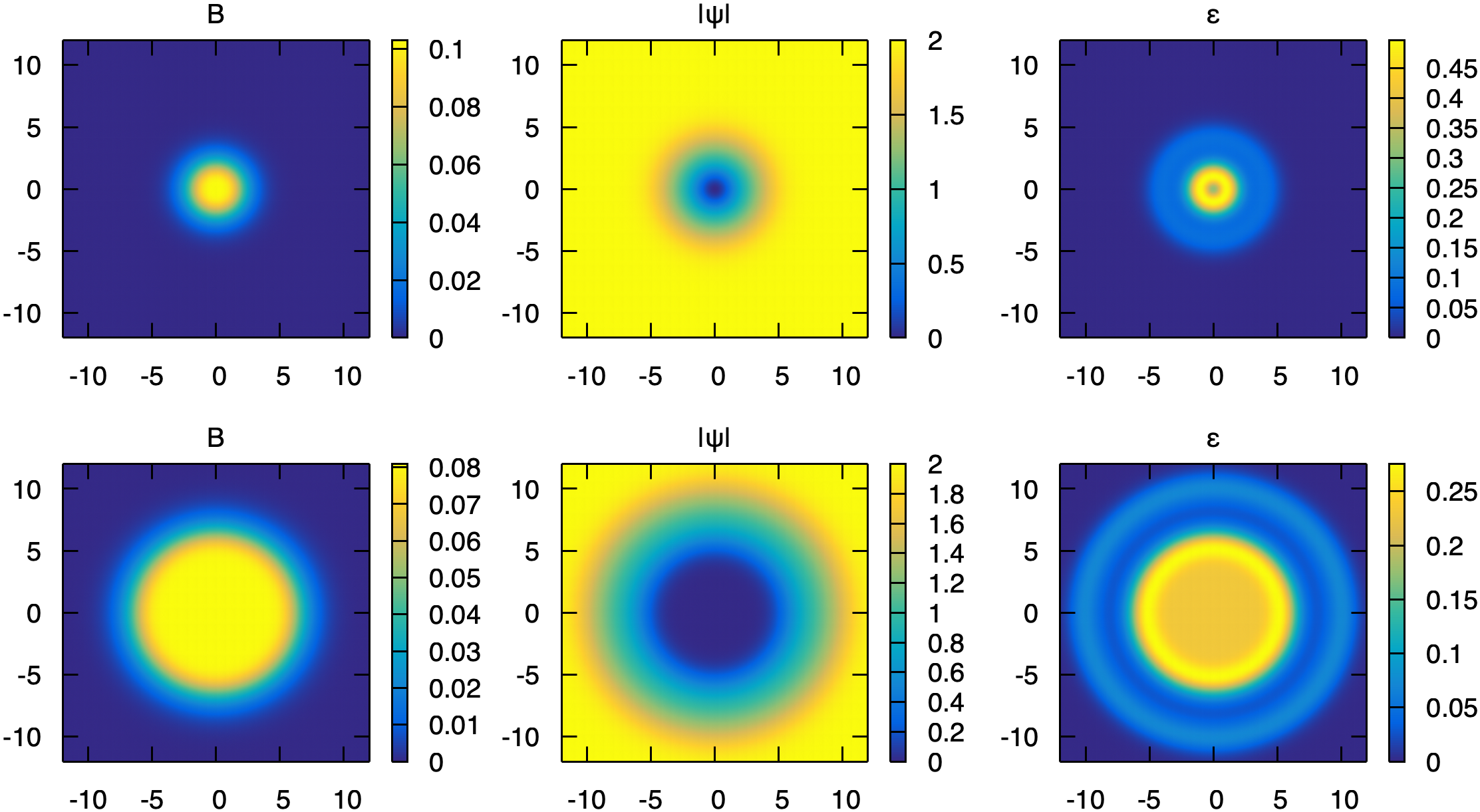}}
  \label{fig:vortexI}
 \caption{Local minimizers of degree $N=2$ and $N=12$ respectively for $\mu^2 = 0.05$, $m_1 = 1$, $m_2 = 2$ and $\alpha = \beta = 0$. The plots include the magnetic field density $B$ the order parameter magnitude $|\psi|$ and the energy density $\mathcal{E}$. These parameters correspond to type I for both vacua, as can be seen in \figref{fig:typology}, hence vortices attract and form a single large winding vortex, contained by the domain wall.}
\end{figure}

For the parameters $\mu^2 = 0.05$ and $\alpha = \beta = 0$ however we are in the type I regime where vortices attract. Hence, we observe that any starting position for the vortices causes them to form local bags which then attract. This results in a single bag with a large winding $N$ vortex contained within it as shown in \figref{fig:vortexI}.

Finally the bags themselves can attract or repel depending on the typology of the outer vacuum $u_2$. Hence we can have four different numerical solutions all of which are shown in \figref{fig:typology}, where the last row shows the result of flowing two well separated $N=2$ bags. The vortices inside the bag can split (repel) or form composite vortices (attract) such that the inner type is type II or type I respectively. The vortex bags can then either attract and form larger bags or repel and form separated bags such that the outer type is type I or type II respectively. This matches the intuition from section 3: the possible solutions are,
\begin{itemize}
\item type I/I - one large bag containing a composite high winding vortex.
\item type II/II - multiple bags containing split single vortices.
\item type II/I - one large bag containing split single vortices.
\item type I/II - multiple bags containing single composite higher winding vortices.
\end{itemize}

\section{Conclusions and future outlook}
We have studied a Ginzburg-Landau or Higgs model with higher order potential term with two distinct vacua. We have demonstrated that the vortex interactions in such a model depend on the superconducting state that the vortices are in. By linearizing around the two vacua we demonstrated 4 different types of system depending on the long-range interactions. The numerical simulations agreed with the predicted long range interactions, leading to our conclusion of 4 different types of vortex interaction:
\begin{itemize}
\item type I/I - one large bag containing a composite high winding vortex.
\item type II/II - multiple bags containing split single vortices.
\item type II/I - one large bag containing split single vortices.
\item type I/II - multiple bags containing single composite higher winding vortices.
\end{itemize}
Note that our analysis only considered long range interactions, it would be instructive to study the short range interactions in such a model \cite{speight2025short}. This has recently been done for vortices with a potential up to 6th order which demonstrated type 1.5 like behavior \cite{sixthOrder}.

We note that when the vortices locally split (inner type II), the resulting vortex clusters look very similar to those observed in type 1.5 systems. In an experimental setting, where the local magnetic field $B$ is measured, there would be no discernible difference between the two as the domain wall "bag" does not couple to the magnetic field and is not visible. 

There are a number of open questions for such a model. While we have proposed this as a simple model for two competing superconducting states, there are many features of a superconductor that will be affected by such a potential choice beyond the vortex interactions. Also the magnetization of a superconductor will now depend on the superconducting Meissner state that it starts in, leading to likely very different magnetization curves.

Finally it has been demonstrated that higher order potentials lead to interesting vortex scattering in Lorentz invariant cosmological systems \cite{martinez2026oscillons}. It would be interesting to study vortex scattering with such a potential. 

\subsection*{Acknowledgements}

This work was supported by the UK Engineering and Physical Sciences Research Council through grant EP/P024688/1, and was commenced while TW was a research fellow at the University of Leeds.

\subsection*{Note added in proof}

During the final preparation of this paper a preprint with substantial overlap was published on the arXiv \cite{Miguelez-Caballero:2026nzv}. 
That paper contains a detailed study of the particular case $\alpha=\beta=0$, including numerical analysis of the spectrum of normal modes about cocentred vortices of winding $N=1,2,3$. The energy minimizers found in \cite{Miguelez-Caballero:2026nzv} agree qualitatively with those reported here in this case ($\alpha=\beta=0$).

\bibliographystyle{stylefile}
\bibliography{bibliography}
\end{document}